\documentclass{article}
\usepackage{arxiv}
\usepackage[utf8]{inputenc} 
\usepackage[T1]{fontenc}    
\usepackage{hyperref}       
\usepackage{url}            
\usepackage{booktabs}       
\usepackage{amsfonts}       
\usepackage{nicefrac}       
\usepackage{microtype}      
\usepackage{lipsum}		
\usepackage{graphicx}
\usepackage{mathrsfs}
\usepackage{doi}
\usepackage{amsmath}
\usepackage{epsfig}
\usepackage[linesnumbered,ruled,vlined]{algorithm2e}
\usepackage{xcolor,soul}

\SetCommentSty{mycommfont}
\usepackage{caption,subcaption}
\allowdisplaybreaks
\usepackage{stmaryrd}
\usepackage{color,soul}
\usepackage{wrapfig}
\usepackage{cite}
\usepackage{cancel}
\hypersetup{
    colorlinks=true,
    linkcolor=blue,
    filecolor=magenta,      
    urlcolor=cyan,
}
\usepackage{courier}

\usepackage{listings}
\usepackage{xcolor}
\definecolor{codegreen}{rgb}{0,0.6,0}
\definecolor{codegray}{rgb}{0.5,0.5,0.5}
\definecolor{codepurple}{rgb}{0.58,0,0.82}
\definecolor{backcolour}{rgb}{0.95,0.95,0.92}

\lstdefinestyle{mystyle}{
  backgroundcolor=\color{backcolour},   commentstyle=\color{codegreen},
  keywordstyle=\color{magenta},
  numberstyle=\tiny\color{codegray},
  stringstyle=\color{codepurple},
  basicstyle=\ttfamily\footnotesize,
  breakatwhitespace=false,         
  breaklines=true,                 
  captionpos=b,                    
  keepspaces=true,                 
  numbers=left,                    
  numbersep=5pt,                  
  showspaces=false,                
  showstringspaces=false,
  showtabs=false,                  
  tabsize=2
}

\title{A fault slip model to study earthquakes due to pore pressure perturbations}

\author{
  {
  Saumik Dana}\\
	University of Southern California\\
	Los Angeles, CA 90007 \\
	\texttt{sdana@usc.edu} \\
 		\And
  {
  Birendra Jha}\\
	University of Southern California\\
	Los Angeles, CA 90007 \\
	\texttt{bjha@usc.edu} 
}

\date{}

\begin{document}
\maketitle
\begin{abstract}
The burgeoning need to sequester anthropogenic CO$_2$ for climate mitigation and the need for energy sustenance leading upto enhanced geothermal energy production has made it incredibly critical to study potential earthquakes due to fluid activity in the subsurface. These earthquakes result from reactivation of faults in the subsurface due to pore pressure perturbations. In this work, we provide a framework to model fault slip due to pore pressure change leading upto quantifying the earthquake magnitude. 
\end{abstract}

\section{Introduction}

\begin{figure}[htb!]
\centering
\includegraphics[scale=0.5]{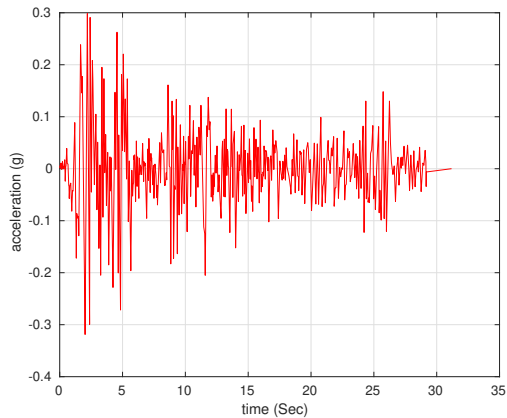}
\caption{1940 El Centro earthquake. Accelerogram (measured as normalized to acceleration due to gravity $g=9.8\,m/s^2$)}
\label{elcentro}
\end{figure}

Earthquakes occur as a result of global plate motion. Some plate boundaries glide past each other smoothly, while others are punctuated by catastrophic failures. Some earthquakes stop after only a few hundred metres while others continue rupturing for a thousand kilometres. The simplest model for earthquake initiation is to assume that when the stress accumulated in the plates exceeds some failure criterion on a fault plane, an earthquake happens \cite{kanamori2004physics}. Evaluating this criterion requires both a measure of the resolved stress on the fault plane and a
quantifiable model for the failure threshold. The groundbreaking work of \cite{anderson1905dynamics} started with the fact that any stress field can be completely described by its principal stresses, which are given by the eigenvectors of the stress tensor and are interpretable as the normal stresses in three orthogonal directions. He then proposed that: (1) the stress state could be resolved by assuming that one principal stress is vertical since the Earth’s surface is a free surface and (2) faulting occurs when the resolved shear stress exceeds the internal friction on some plane in the medium. Internal friction is defined analogously with conventional sliding friction as a shear stress proportional to the normal stress on a plane. One complication to this simple picture was recognized early on. High fluid pressures can support part of the load across a fault and reduce the friction. The importance of the fluid effect on fault friction was first recognized by \cite{king1959role}. In the course of their work on oil exploration, they observed that pressures in pockets of fluids in the crust commonly exceeded hydrostatic pressure. They connected this observation with studies of faulting and proposed that the pore pressure at a depth can approach the normal stress on faults, resulting in low friction. Fig. \ref{elcentro} shows the seismograph recorded acceleration of the 1940 EL Centro earthquake that occurred in the Imperial Valley in southeastern Southern California near the USA-Mexico border. It was the first major earthquake to be recorded by a strong-motion seismograph located next to a fault rupture, and led to a total damage of $\$6$ million \cite{stover1993seismicity}. 

\subsection{Enhanced geothermal systems (EGS)}

\begin{figure}[htb!]
\begin{subfigure}{.45\textwidth}
\centering
\includegraphics[scale=0.30]{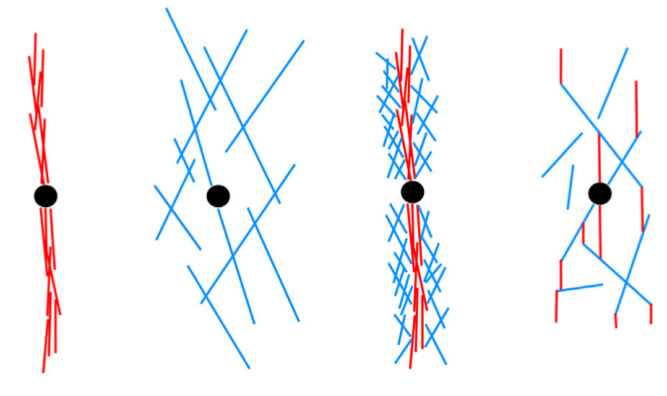}
\caption{Four conceptual models for EGS stimulation. The black dot represents the wellbore. New fractures are represented with red
lines, and preexisting fractures are represented with blue lines. Source:~\cite{mcclure2014investigation}}
\label{mcclure}
\end{subfigure}
\hspace{20pt}
\begin{subfigure}{.45\textwidth}
\centering
\includegraphics[scale=0.25]{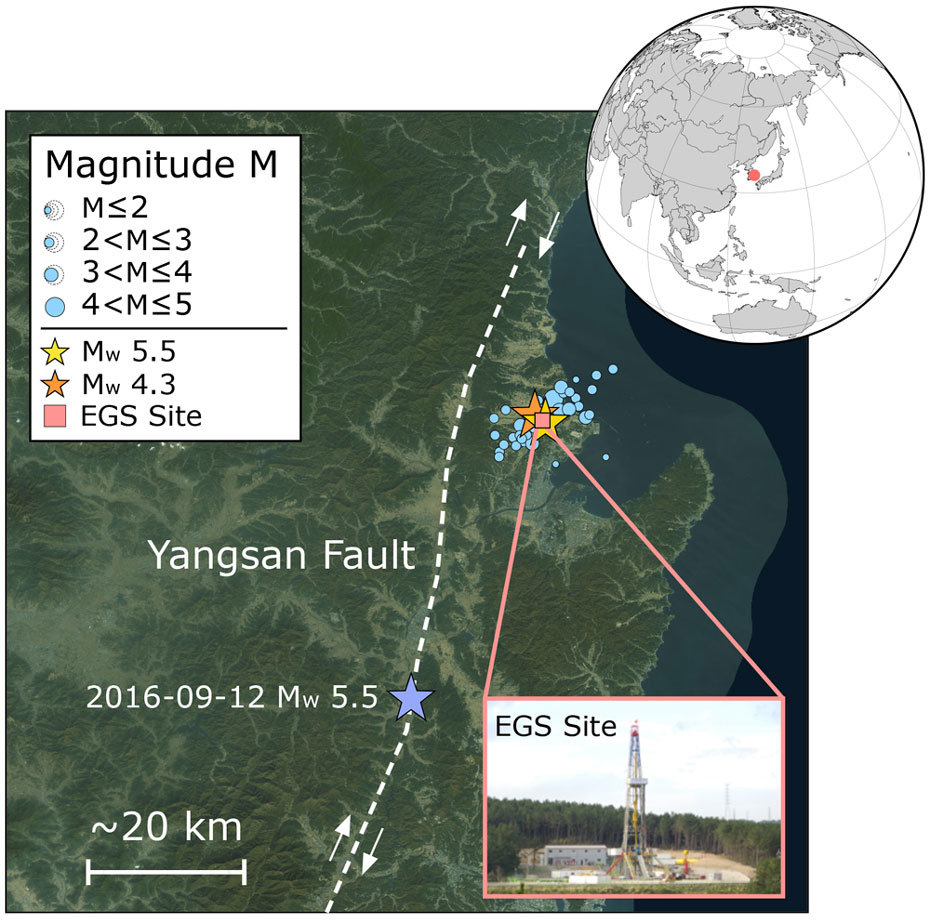}
\caption{Seismicity in the Pohang EGS site. Source:~\cite{grigoli2018november}}
\label{pohang}
\end{subfigure}
\caption{EGS}
\end{figure}

Pulse fracturing in hot dry rock is a complex multi-physics problem which involves nucleation and propagation of cracks by injecting a combination of padding fluid/proppant/propellant at high pressure down a wellbore~\cite{hubbert1957mechanics,montgomery2010hydraulic,king2010thirty,adams2013differentiating,osiptsov2017fluid,tariq2019review,white2014modeling,PARCHEIESFAHANI2020107152,zang2017keynote,wang2020parameters}, and has documented long lasting implications for EGS~\cite{king2010thirty,mcclure2014investigation,li2015analysis,kutun2018hydraulic,zang2019reduce}. As shown in Fig.~\ref{mcclure}, the conceptual models that have been proposed in literature as stimulation mechanisms in EGS~\cite{mcclure2014investigation} include (1) Only new, propagating fractures contribute to permeability enhancement, (2) Stimulation occurs only through induced slip on preexisting fractures, (3) Continuous new fractures propagate away from the wellbore, but fluid leaks off into natural fractures, which slip and experience enhanced transmissivity, and (4) Continuous pathways for flow involve both new and preexisting fractures. The fourth mechanism has been widely accepted to model hydraulic stimulation of EGS~\cite{kamali2018analysis,norbeck2018field,norbeck2018exploring,abe2019investigating}. The fracture initiated from a sheared preexisting natural fracture is called a wing crack or a splay fracture. Wing cracks found in rock are tensile fractures but they are different from hydraulic fractures because (1) wing cracks are initiated from the tension field induced by shear slip of a preexisting natural fracture while hydraulic fractures are initiated from an injection well by fluid pressure; (2) wing cracks are curved cracks while hydraulic fractures propagate straight and perpendicularly to the least principal stress; and (3) tension forces to open wing cracks are supported both by fluid pressure and shear slip of a preexisting natural fracture while tension forces to open hydraulic fractures are supported only by fluid pressure~\cite{abe2019investigating}. Hitherto, HF has been associated with microseismic events due to their small magnitudes. However, there is now growing evidence to suggest that HF induced fault activity in EGS projects across the world has caused larger earthquakes~\cite{grigoli2018november,kamali2018analysis,norbeck2018exploring}. Fig.~\ref{pohang} shows the distribution of seismicity in the Pohang, South Korea EGS site. Postulates suggest the actual amount of injected fluid was only $0.2\%$ of what is theoretically expected for the largest recorded earthquake at the site~\cite{grigoli2018november,schultz2020}.
The causation is established via spatio-temporal correlation; the events occur close ($<10$ km) to the injection well and shortly (minutes to hours) after injection. Most of the sites are in geographical regions with little to no history of prior natural seismicity, which further supports the causative link between fracturing and the observed seismic events. 

\subsection{Earthquake triggering mechanisms}

\begin{figure}[htb!]
\centering
\includegraphics[scale=0.175]{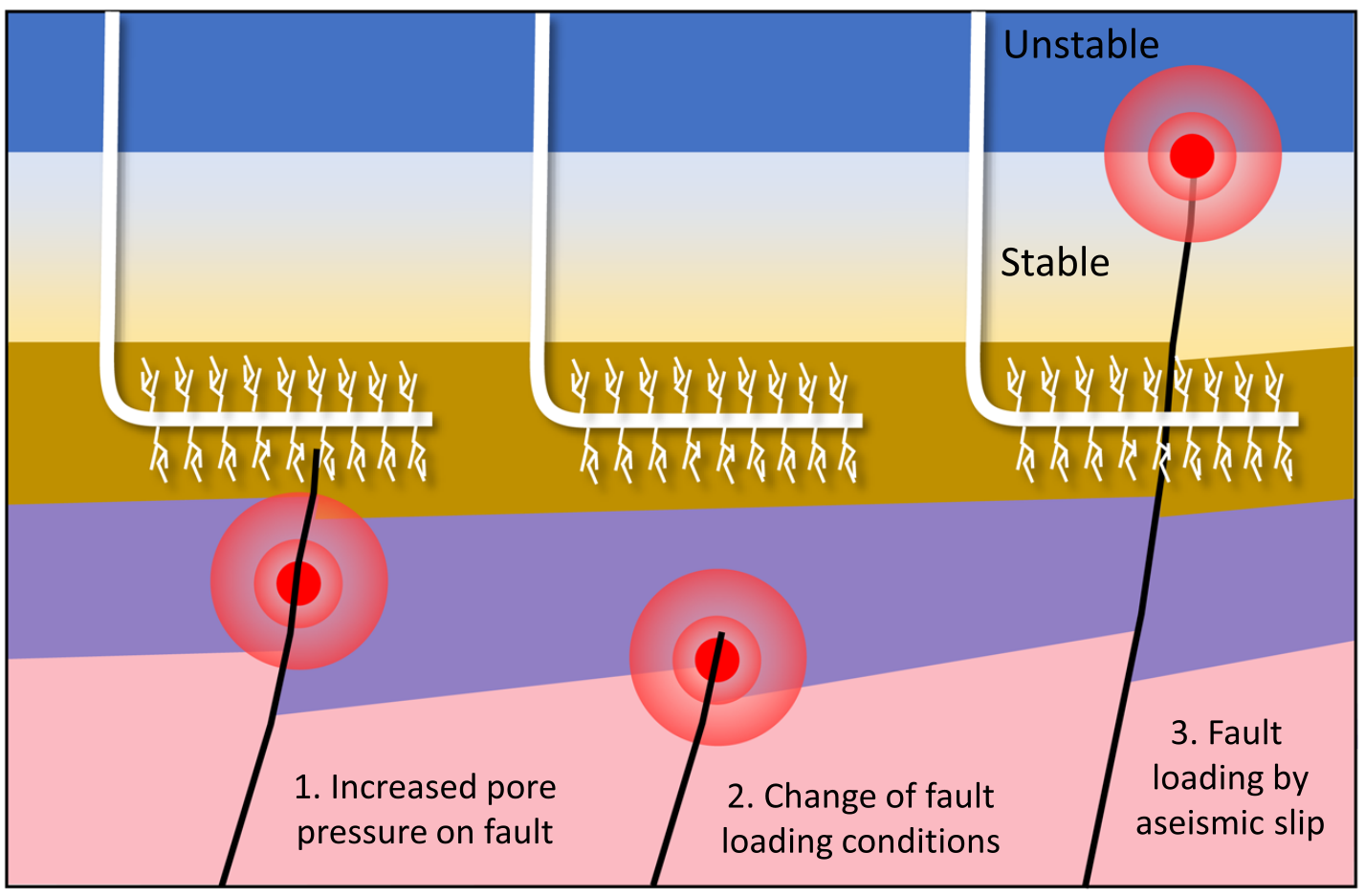}
\caption{Three proposed earthquake triggering mechanisms in literature. 
Source: \cite{schultz2020}}
\label{hf_seismicity}
\end{figure}

The earthquake triggering mechanisms (see Fig.~\ref{hf_seismicity}) that have been proposed are: (1) propagation of the fluid pressure diffusion front through the newly created fracture network, followed by pressurization/lubrication of a basement fault that slips seismically~\cite{ShaS2003}, (2) poroelastic stress transfer from slipped fractures on to the fault causing an increase in shear and/or tension on the fault followed by seismic slip, and (3) fluid injection-induced aseismic rupture front which propagates faster and to larger distances than the fluid pressure diffusion front, thereby transmitting stresses to unstable portions of the fault beyond the fluid-pressurized region~\cite{BhaP2019}. All three mechanisms can activate a critically stressed fault (initial stress state close to frictional failure), which can culminate into an induced seismic event~\cite{foulger2018global,schultz2020,atkinson2020, davies,maxwell2013unintentional,clark2012hydraulic,westwood2017horizontal,keranen2018induced,krupnick2017induced}.
In quasi-static simulations, which neglect elastodynamics and wave propagation effects, the induced seismicity mechanisms are often quantified using Coulomb Failure Function (CFF)~\cite{ReaP1992,jha2014coupled,ZhaX2019}, which includes contributions from the effective stress and fault's friction coefficient. The effective stress is defined as the difference between the total stress (from tectonics and expansion/contraction of surrounding rock) and the pore pressure (from injection) scaled by the Biot coefficient. The friction coefficient evolution often requires a fault rheological model e.g., the Slip Weakening or the Rate- and State-dependent Friction (RSF) model. RSF is the gold standard for modeling earthquake cycles (inter-seismic loading followed by co-seismic relaxation) on  faults~\cite{DieJ1979,DieJ1981,RicJ2001}, whereas slip weakening provides a reasonable approximation for modeling the drop in friction during a single seismic/aseismic event. CFF is computed using the fault traction vector which is given by the product of the effective stress tensor and the fault normal vector. Initially, when the fault is locked and under equilibrium with the tectonic and hydrostatic stresses, $\textrm{CFF}<0$.  As injection proceeds, the pore pressure increases and the near-wellbore region expands volumetrically, which applies compression on the region farther from the wellbore. The spatial distribution of the total stress tensor and pressure around the fault change. An increase in CFF, i.e. $\Delta \textrm{CFF}>0$, indicates fluid-induced destabilization of the fault. 
When CFF exceeds fault's intrinsic cohesion, the fault may slip. The slip releases the host rock strain energy, accumulated from natural and/or anthropogenic loading, over multiple years. The slip can be seismic or aseismic depending on fault's frictional properties (e.g., velocity-weakening vs. velocity-strengthening) and the fault loading rate. Similarly, a decrease in fault's CFF, e.g., due to stresses transferred from a newly created HF, can induce mechanical stabilization of the fault.

\section{Fault slip model}

Elastodynamics~\cite{ThomasLapusta2014,JinL2018} and quasi-dynamic~\cite{Mcclure2012,Pamillon2018} approaches allow calculation of the seismic magnitude from dynamic fault rupture simulations~\cite{aniko20213d,lindsey2021slip,ragon2021accounting,allison2021earthquake,lui2021role,bai2021displacement}. Elastodynamics includes the inertia term in the equilibrium equation, whereas quasi-dynamic approximates inertial effects via a radiation damping term~\cite{RicJ1993}. 
\subsection{Displacement discontinuity}
\begin{figure}[htb!]
\begin{subfigure}{.3\textwidth}
\centering
\includegraphics[trim={0 0 0 0},clip,scale=0.75]{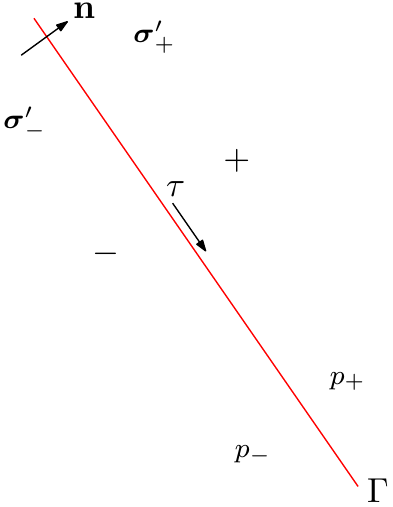}
\caption{The two sides of the fault surface, which need not be planar, are designated as the `$+$' side and the `$-$' side. The fault normal vector $\mathbf{n}$ points from the negative side to the positive side of the fault. $\boldsymbol{\sigma}'$ is the effective stress tensor.}
\label{sketch10}
\end{subfigure}
\hspace{20pt}
\begin{subfigure}{.6\textwidth}
\centering
\includegraphics[trim={0 0 0 0},clip,scale=0.75]{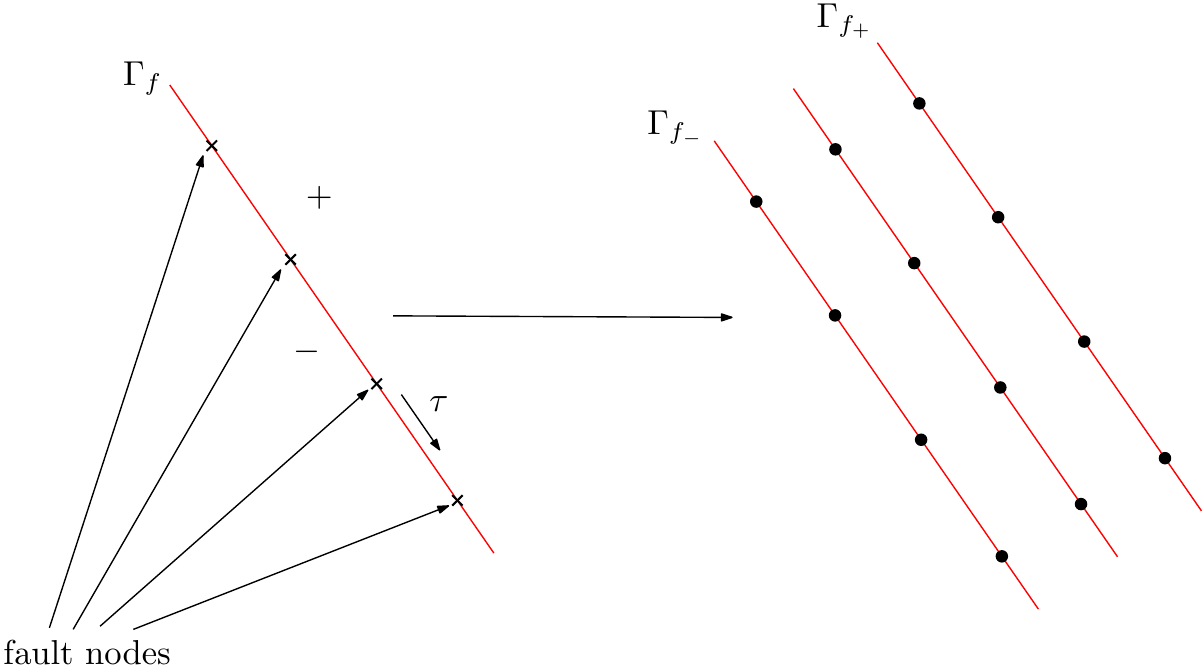}
\caption{Fault nodes are triplicated to solve for Lagrange multipliers on the fault nodes and displacements on either side of the fault nodes.}
\label{sketch11}
\end{subfigure}
\caption{Fault mechanics}
\end{figure}
As shown in Figs.~\ref{sketch10} and ~\ref{sketch11}, we treat faults as surfaces of discontinuity embedded in the continuum across which displacement is allowed to be discontinuous to recognize the possibility of fault slip \cite{jha2014coupled,aagaard2}. Slip on the fault is the displacement of the positive side relative to the negative side,
\begin{equation}\label{e:slip}
(\boldsymbol{u}_+ - \boldsymbol{u}_-)-\boldsymbol{d}=\boldsymbol{0} \; \text{on} \; \Gamma_f,
\end{equation}
where $\boldsymbol{d}$ is the fault slip vector. We impose the effective traction on the fault by introducing a Lagrange multiplier, $\boldsymbol{l}$, which is a force per unit area required to satisfy Eq. \eqref{e:slip}. A positive value of $\boldsymbol{l}\cdot\boldsymbol{n}$ indicates that a tensile effective stress is transmitted across the fault surface. The Kuhn-Tucker conditions of contact mechanics are obeyed such that no penetration occurs and the effective normal traction stays compressive at the contact surface. 
\subsection{Fault traction}
The shear traction on the fault is computed as
\begin{align}
\label{fault_traction}
\tau = \vert \mathbf{l} - (\mathbf{l}\cdot \mathbf{n}) \mathbf{n} \vert
\end{align}
We use the Mohr-Coulomb theory to define the stability criterion for the fault~\cite{JaeJ1979}. When the shear traction on the fault is below the friction stress, $\tau \le \tau_f$, the fault does not slip. When the shear traction is larger than the friction stress, $\tau>\tau_f$, the contact problem is solved to determine the Lagrange multipliers and slip on the fault, such that the Lagrange multipliers are compatible with the frictional stress \cite{jha2014coupled}. 
\subsection{Fault friction}
The frictional stress $\tau_f$ on the fault as 
\begin{equation}\label{e:friction}
\tau_f = 
\begin{cases} 
\tau_c - \mu \boldsymbol{l}\cdot\boldsymbol{n},  &\boldsymbol{l}\cdot\boldsymbol{n} < 0, \\
\tau_c,  &\boldsymbol{l}\cdot\boldsymbol{n}\ge 0, \end{cases}
\end{equation}
where $\tau_c$ is the cohesive strength of the fault and $\mu$ is the coefficient of friction. The rate- and state-dependent friction model~\cite{DieJ1979,DieJ1981,RuiA1983,SchC1989,MarC1998} is
\begin{equation}\label{e:ratestate}
\begin{split}
\mu &= \mu_0 + A\ln{\left(\frac{V}{V_0}\right)} + B\ln{\left(\frac{V_0\theta}{d_c}\right)}, \\
\frac{d\theta}{dt}&= 1-\frac{\theta V}{d_c},
\end{split}
\end{equation}
where $V=|d\boldsymbol{d}/dt|$ is the slip rate magnitude, $\mu_0$ is the steady-state friction coefficient at the reference slip rate $V_0$, $A$ and $B$ are empirical dimensionless constants, $\theta$ is the macroscopic variable characterizing state of the surface and $d_c$ is a critical slip distance. Here, $\theta$ may be understood as the frictional contact time \cite{DieJ1979}, or the average maturity of contact asperities between the sliding surfaces \cite{RicJ1993}. The evolution of $\theta$ is assumed to be independent of changes in the normal traction that can accompany the fault slip due to changes in fluid pressure. The model accounts for the decrease in friction (slip-weakening) as the slip increases, and the increase in friction (healing) as the time of contact or slip velocity increase. The two effects act together such that $A>B$ leads to strengthening of the fault, stable sliding and creeping motion, and $A<B$ leads to weakening of the fault, frictional instability, and accelerating slip. In this way, the model is capable of capturing repetitive stick-slip behavior of faults and the resulting seismic cycle \cite{DieJ1981, SchC1989}. 
\subsection{Fault pressure}
A difference in fluid pressure across the fault leads to a pressure jump $\llbracket p\rrbracket_{\Gamma}=p_+-p_-$, where $p_+$ and $p_-$ are the equivalent multiphase pressures on the `positive' and the `negative' side of the fault. This pressure jump leads to a discontinuity in the effective stress across the fault, such that the total stress is continuous,
\begin{equation}
 \boldsymbol{\sigma}'_-\cdot\boldsymbol{n}-bp_-\boldsymbol{n}=\boldsymbol{\sigma}'_+\cdot\boldsymbol{n}-bp_+\boldsymbol{n}, 
\end{equation} 
where $b$ is the Biot coefficient. Fault stability can be assessed by evaluating the stability criterion on both sides of the fault separately. The side of the fault where the criterion is met first determines the fault stability. Equivalently, this can be achieved by defining a \emph{fault pressure} that is a function of the pressures on the two sides, $p_+$ and $p_-$. Introducing the fault pressure allows us to uniquely define the \emph{effective} normal traction on the fault, $\sigma'_n$, and determine the fault friction $\tau_f$ (Eq.~\eqref{e:friction}). Since the stability criterion $\tau > \tau_f$ is first violated with the larger pressure, we define the fault pressure $p$ as
\begin{equation}\label{e:faultpres}
p=\max{(p_-,  p_+)}.
\end{equation}
\subsection{Earthquake magnitude}
In case of the elastodynamics framework with the RSF fault model, once the slip is nucleated at the hypocenter, wave propagation is simulated during seismic slip. Depending on the hypocenter location, the natural fractures, HF, distant portions of the fault, and reservoir boundaries can act as barriers to the seismic wave fronts.  Such barriers can be treated as inclusions using absorbing boundary layers (ABLs)~\cite{Sun2016} or perfectly matched layers (PMLs)~\cite{Berenger_1994,Basu_Chopra_2003}. The moment-based seismic magnitude~\cite{HanT1979} is computed as $M_w=\frac{2}{3}\log_{10}M_0-6.0$, where the seismic moment is $M_0=\int_{\Gamma_f} G |\boldsymbol{d}| d\Gamma$, $|\boldsymbol{d}|$ is the magnitude of the slip vector (accumulated over the event duration). 
\section{Numerical challenges}
\begin{figure}[htb!]
\centering
\includegraphics[scale=0.80]{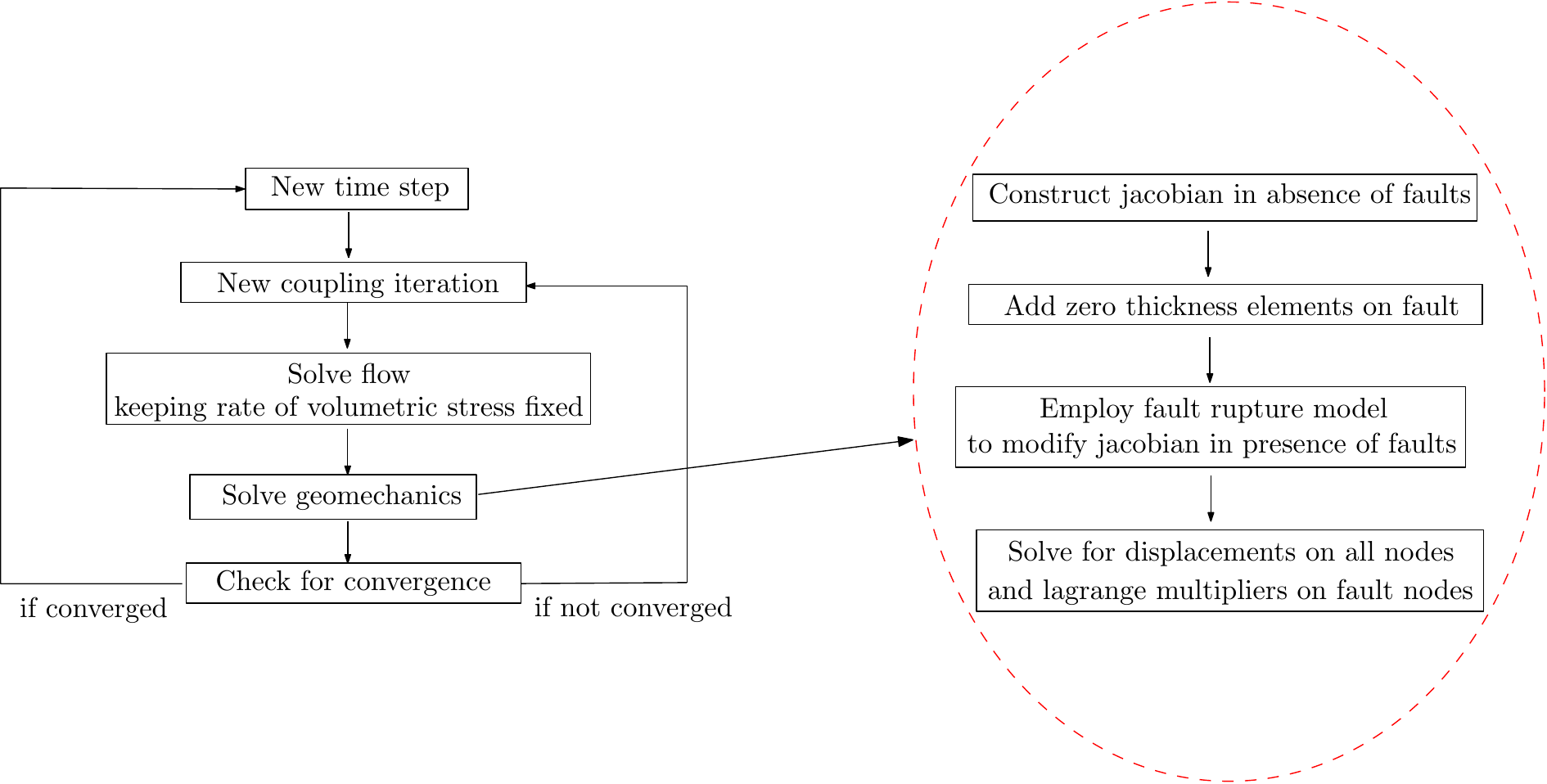}
\caption{Algorithmic flowchart}
\label{faultsintogeo}
\end{figure}

As shown in Fig.~\ref{faultsintogeo}, we deploy the fixed stress split algorithm~\cite{dana-2018,dana2019design,dana2019system,dana2020,dana2021,danacg,danacmame,danathesis,JAMMOUL2020,jammoul2019RSC} in the open source geodynamics framework~\cite{aagaard1,aagaard2} coupled with a multiphase flow code~\cite{jha2014coupled} to solve the coupled problem. While it is well-known that the multiphase flow problem brings in a host of numerical issues associated with the simultaneous resolution of pressure-saturation primary variables~\cite{rasmussen2021open,jiang2021smooth,nasir2021two,li2021sequential}, the presence of faults renders the geomechanics as a saddle point problem. Although the geodynamics framework handles the saddle point problem efficiently, the coupling with the multiphase flow code brings in a set of challenges that require careful consideration of the solvers deployed for both flow and geomechanics. In addition to that, the elastodynamics brings in time stepping issues associated with the explicit update of the displacement primary variable. We shall tackle all these challenges in future work.   

\appendix

\bibliographystyle{unsrt}     
\bibliography{sample}   
\end{document}